\documentclass[12pt,letterpaper]{article}

\usepackage{amsmath}
\usepackage{amsfonts}
\usepackage{amssymb}
\usepackage{graphicx}
\usepackage{epsfig}
\usepackage{epstopdf}
\usepackage{color}
\usepackage{multirow}

\def\be{\begin{equation}}
\def\ee{\end{equation}}

\newcommand{\fnlloc}{f_{NL}^{local}}

\def\be{\begin{equation}}
\def\ee{\end{equation}}
\def\ba{\begin{eqnarray}}
\def\ea{\end{eqnarray}}
\def\nn{\nonumber}

\def\threej#1#2#3#4#5#6{\left( \begin{array}{ccc} #1 & #2 & #3 \\ #4 & #5 & #6 \end{array} \right) }

\def\Npix{N_{\rm pix}}
\def\Nalm{N_{\rm alm}}

\def\Nchan{N_{\rm chan}}

\def\ellmin{\ell_{\rm min}}
\def\ellmax{\ell_{\rm max}}

\def\fnlloc{f_{NL}^{\rm local}}

\def\fnleq{f_{NL}^{\rm equil}}

\def\ha{\hat a}
\def\hs{\hat s}

\def\n{\widehat{\bf n}}
\def\hfnl{{\widehat f_{NL}}}

\begin{document}

\begin{flushright} {\footnotesize HUTP-0X/XXXXX} \end{flushright}
\vspace{5mm}
\vspace{0.5cm}
\begin{center}

\def\thefootnote{\fnsymbol{footnote}}

{\Large \bf Optimal limits on $\fnlloc$ from WMAP 5-year data}  \\[1cm]
{\large Kendrick M. Smith$^{\rm a}$, Leonardo Senatore$^{\rm b,c,d}$ and Matias Zaldarriaga$^{\rm c,d}$}
\\[0.5cm]

\vspace{.2cm}

{\small 
\textit{$^{\rm a}$ Institute of Astronomy, \\
Cambridge University, Cambridge, CB3 0HA, UK}} 

\vspace{.2cm}

{\small 
\textit{$^{\rm b}$ School of Natural Sciences, Institute for Advanced Study, \\
Olden Lane, Princeton, NJ 08540, USA
}}

\vspace{.2cm}

{\small 
\textit{$^{\rm c}$ Center for Astrophysics, \\
Harvard University, Cambridge, MA 02138, USA
}}

\vspace{.2cm}

{\small 
\textit{$^{\rm d}$ Jefferson Physical Laboratory, \\
Harvard University, Cambridge, MA 02138, USA
}}
\end{center}

\vspace{.8cm}

\hrule \vspace{0.3cm} 
{\small  \noindent \textbf{Abstract} \\[0.3cm]
\noindent
We have applied the optimal estimator for $\fnlloc$ to the 5 year WMAP data. Marginalizing over the amplitude of foreground templates we get 
$-4< \fnlloc < 80$ at 95\% CL. Error bars of previous (sub-optimal) analyses are roughly 40\% larger than these. The probability that a Gaussian simulation, analyzed using our estimator, gives a result larger in magnitude than the one we find is 7\,\%. Our pipeline gives consistent results when applied to the three and five year WMAP data releases and agrees well with the results from our own sub-optimal pipeline. We find no evidence of any residual foreground contamination.}
   
\vspace{0.5cm}  \hrule

\def\thefootnote{\arabic{footnote}}
\setcounter{footnote}{0}

\section{Introduction}

It has become apparent that departures from Gaussianity of the primordial perturbations could shed light on the physics of inflation. In most inflationary models perturbations tend to be very close to Gaussian with possible observable departures only for the three-point function (or bispectrum). Single field models of inflation, in which the quantum fluctuations of the same field that dominates the energy density during inflation become the seeds for structure formation, satisfy a consistency relation that relates the shape of the three-point function to the dynamics of the inflaton field (eg. \cite{Maldacena:2002vr,Creminelli:2004yq,Cheung:2007sv}). These models produce a bispectrum that is either un-measurably small or, when large, of the so-called equilateral shape whose amplitude is usually denoted by $\fnleq$ \cite{Babich:2004gb,Cheung:2007st}. Testing the inflationary bispectrum could allow us to distinguish single field models from other alternatives, or  measure the ``sound-speed'' ($c_s$) of perturbations during inflation \cite{Cheung:2007st,Chen:2006nt}. 
Models where  the fluctuations of a field other than the inflation seed the observed large scale structure, as  it happens for example in some multi-field inflationary models \cite{Lyth:2002my,Zaldarriaga:2003my}, fall in a different class. These models produce a bispectrum of the so-called local type, whose amplitude peaks in the squeezed limit. This bispectrum encodes correlations between modes that exited the horizon at very different times during inflation~\footnote{The same mechanism can be effective during the contracting phase of the new Ekpyrotic universe as well \cite{Creminelli:2007aq}.}. Large correlations of this type are forbidden in single field models. The amplitude of the bispectrum of the local type is usually parametrized by $\fnlloc$. 

\begin{figure}[!ht]
\centerline{\epsfxsize=12cm\epsffile{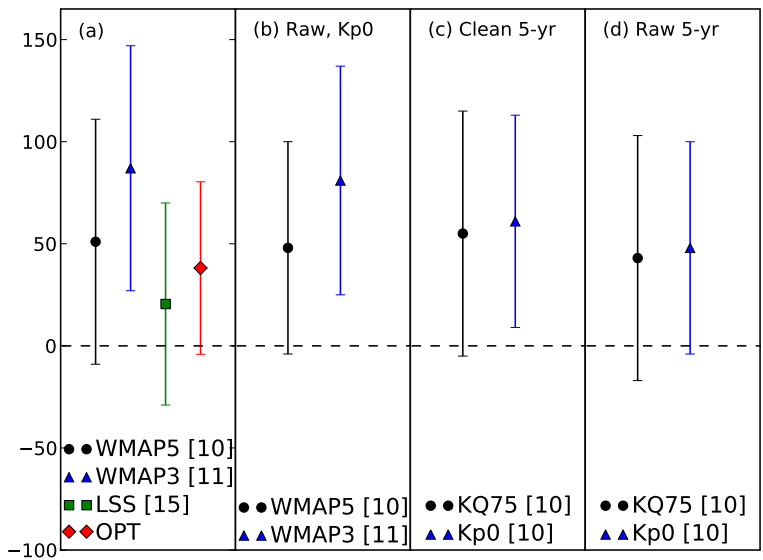}}
\caption{Current constraints on $\fnlloc$. Errors in this figure and throughout the paper are 2-$\sigma$. Panel (a) best results from WMAP 5 years from the WMAP team \cite{Komatsu:2008hk} and WMAP 3 years from Yadav \& Wandelt \cite{Yadav:2007yy} together with the large scale structure results from Slosar et al \cite{Slosar:2008hx} and the results from this paper using our optimal method (OPT). Panel (b) comparison of  \cite{Komatsu:2008hk} and \cite{Yadav:2007yy} for the same choice of analysis parameters ($l_{max}=500$,  raw maps and the Kp0 mask). Panels (c) and (d) show the effect of the mask for cleaned and raw maps respectively (from \cite{Komatsu:2008hk}). }
\label{status}
\end{figure}

The most recent search for non-gaussianity in  CMB data was done by the WMAP team which used 5 years worth of data to put constraints on both the local and equilateral shapes. Their best estimates are: $-9 < \fnlloc < 111$  (95\% CL) and  $-151 < \fnleq < 253$ (95\% CL), both consistent with Gaussian initial conditions \cite{Komatsu:2008hk}. These results are not without puzzles.  Panel (a) of figure \ref{status} compares the quoted WMAP results to those obtained earlier by a different group \cite{Yadav:2007yy}, where a detection of non-gaussianity of the local type was claimed: $27< \fnlloc < 147$  (95\% CL).

A detection of local non-gaussianity would have profound consequences for our understanding of Inflation, ruling out all single field inflation models. Thus it is important to understand what changed. The error bars in figure \ref{status} are dominated by the cosmic variance of the large scale modes so the shift seen between 3 and 5 years was not expected. Furthermore both analyses used basically the same method to constrain $\fnlloc$.  

Indeed if one compares \cite{Komatsu:2008hk} and \cite{Yadav:2007yy} for the same choice of analysis parameters ($\ellmax=500$, using raw maps and the Kp0 mask), the shift in results is remarkable (panel (b) of figure \ref{status}). While  \cite{Komatsu:2008hk} gets $-4< \fnlloc < 100$, \cite{Yadav:2007yy} gets $25< \fnlloc < 135$  (95\% CL).  Notice that indeed the size of the 95\% confidence interval  has changed little, meaning that there is not that much additional information in the 5 year data set. The error bars are dominated by the ``cosmic-variance'' component which is common to both data sets as they are observing the same sky. The shift of the mean value between both analysis was dramatic (however, one should be careful when comparing these results, as they were obtained by different
groups with slightly different ways of weighting the data). Did something change in the data? 

A natural worry when searching for deviations from Gaussianity is the effect of foregrounds. The WMAP team in their 5 year papers advocates using foreground cleaned maps and a new mask, the KQ75 mask, which is larger than the Kp0 mask (the standard mask used in analysis of the 3 year data release). The KQ75 mask cuts out 4.9\% more sky then Kp0 ($f_{sky}^{Kp0}=0.765$ and $f_{sky}^{KQ75}=0.716$). The authors of \cite{Yadav:2007yy}  on the other hand advocated looking at raw maps arguing that foregrounds appear to bias estimates of $\fnlloc$ negative (as discussed later we do not agree with this conclusion). If foregrounds bias estimates negative then using raw maps results in a lower limit for  $\fnlloc$ which was found to be positive. It would then seem that using raw maps only strengthens the significance of the detection in \cite{Yadav:2007yy}. 

Panel (c) of figure \ref{status} compares the effect of changing the mask when analyzing clean maps. Panel (d) of figure \ref{status} compares the effect of changing the mask when analyzing raw maps. Both sets of results are from \cite{Komatsu:2008hk}.  The first thing to note is that the choice of mask makes a difference, shifting for example the range from $-4< \fnlloc < 100$ for raw maps with Kp0 to $-17< \fnlloc < 103$ for raw maps with KQ75. Notice that the 95 \% range increased by 15\%, much more than the expected 3.5\% increase that results from a  $\sqrt{f_{sky}}$ scaling of the error bars.  Using cleaned vs raw maps had a more dramatic effect on the mean value of $\fnlloc$. In the case of the Kp0 mask,  the range changed from $-4< \fnlloc < 100$ for raw maps  to $9< \fnlloc < 113$ which exceeds zero at 95\% CL. The excess is not signficant at 95\% CL in the cleaned maps masked with KQ75 ($-5< \fnlloc < 115$). Note that the increase in the error bars as one moves from Kp0 to KQ75 is in large part responsible for the decreased statistical significance of the excess. This increase of the error bars is directly related to the lack of optimality of the old algorithm, so the situation is not fully satisfactory. 

In summary, there was a large shift between the 5 and 3 year results of \cite{Komatsu:2008hk} and \cite{Yadav:2007yy}. Furthermore even within 5 year results masked with KQ75 foregrounds are still somewhat of an issue in that they change the results when comparing raw and cleaned maps. Thus to correctly asses the significance of a non-gaussianity detection it would be preferable not just to use clean maps but to be more conservative and marginalize over foregrounds including the uncertainty in the foreground cleaning procedure into the final error bars. Finally the analysis method seems too sensitive to the choice of mask and that sensitivity accounts for part of the decrease in the significance.  

Given the importance of a detection of local non-Gaussinity it is imperative to improve the situation. We will do so by analyzing the data using the optimal estimator found in \cite{Creminelli:2005hu}, with the implementation developed in \cite{Smith:2006ud}, and by improving the treatment of foregrounds. Using the standard estimator results in error bars that are 40\% larger than those obtained here.  Our best estimate is $-4< \fnlloc < 80$ at 95\% CL.
 
There is another probe of non-gaussianity that can compete with the CMB in terms of statistical power, the measurement of the scale dependence of the bias of large scale structure tracers \cite{Dalal:2007cu}. The first result obtained using this technique  $-29< \fnlloc < 70$ is consistent with gaussianity (panel (a) of figure \ref{status}) and has error similar to those obtained with the CMB \cite{Slosar:2008hx}. It may be early days for this new probe, but current results at least disfavor a large $\fnlloc$.

If we combine the optimal WMAP5 result from this paper with the SDSS result from \cite{Slosar:2008hx}, we get $-1 < \fnlloc < 63$ at 95\% CL.
Constraints on $\fnlloc$ from other datasets currently have negligible statistical weight in comparison to WMAP5+SDSS, so this last result
combines essentially all the data to date.

In section \ref{methods} we will summarize our methods, in section \ref{results} we present our results,  in sections \ref{tests1} and \ref{tests2} we describe tests of the robustness of our results and we conclude in \ref{conclusions}.  We leave some technical details to the appendix. 

\vspace{0.3cm}

\section{Summary of analysis methods}\label{methods}

\subsection{Optimal analysis}

The optimal (i.e. minimum-variance) estimator for an arbitrary bispectrum $B_{\ell_1\ell_2\ell_3}$ was constructed in
\cite{Creminelli:2005hu,Creminelli:2006gc}, building on previous work in \cite{Komatsu:2003iq}, and shown to contain both cubic and linear terms:
\be
\widehat {\mathcal E} = \frac{1}{\mathcal N} B_{\ell_1\ell_2\ell_3} \threej{\ell_1}{\ell_2}{\ell_3}{m_1}{m_2}{m_3}
       \Big[ (C^{-1}\ha)_{\ell_1 m_1} (C^{-1}\ha)_{\ell_2 m_2} (C^{-1}\ha)_{\ell_3 m_3} 
         - 3 C^{-1}_{\ell_1 m_1, \ell_2 m_2} (C^{-1}\ha)_{\ell_3 m_3} \Big] \label{eq:Edef}
\ee
where ${\mathcal N}$ is a constant which normalizes the estimator to have unit response to $B_{\ell_1\ell_2\ell_3}$.
Here, $\ha_{\ell m}$ is assumed to be a noisy measurement of the CMB with signal + noise covariance $C=(S+N)$.
The $C^{-1}$ filter appearing in Eq.~(\ref{eq:Edef}) optimally weights the data in the presence of complications such as
multiple data channels (with different beams), inhomogeneous noise, the sky cut, or modes of the data which we want to
marginalize such as the monopole and dipole~\footnote{In reality, as shown in \cite{Creminelli:2006gc} and verified numerically in \cite{Liguori:2007sj}, in the case of a significant detection of $f_{NL}^{\rm local}$, the estimator in eq.~(1) becomes suboptimal, and a simple correction to the normalization has to be implemented to make it optimal again \cite{Creminelli:2006gc}. For the central value of $f_{NL}^{\rm local}$ that we will find from our analysis of the WMAP 5 yr data, this effect is quite irrelevant, affecting our error bars at the order of 10\,\%, and we decide to neglect it. In particular, if we just want to determine if and at what statistical level a zero value of $f_{NL}^{\rm local}$ is excluded, than the result of our estimator when applied to the data has to be compared against Gaussian simulations. In this case, our estimator is always optimal. Since no significant detection of $f_{NL}^{\rm local}$ has been made so far, this will be the approach taken in this paper.}.

Current estimates of $\fnlloc$ from WMAP data \cite{Komatsu:2008hk,Yadav:2007yy} do not use the optimal estimator, 
due to the implementational difficulty and CPU cost of the $C^{-1}$ operation.
Instead, a suboptimal estimator is defined by replacing $C^{-1}$ in Eq.~(\ref{eq:Edef}) by a heuristically constructed filter.
The error on $\fnlloc$ using the suboptimal estimator is about 40\% larger than the optimal value.
In this paper, we will present the first optimal analysis of WMAP data.
The details of our $C^{-1}$ implementation are taken from \cite{Smith:2007rg} and summarized in Appendix~\ref{app:estimator}.

In additon to reducing the error, the optimal estimator also has the advantage of eliminating {\em a posteriori} choices that can
introduce bias or complicate interpretation of the results.
This is particularly important in WMAP, where the evidence for nonzero $\fnlloc$ currently has borderline statistical significance,
and is sensitive to the choice of $\ellmax$.
The suboptimal estimator is not unique: different implementations can make different ``arbitrary'' choices of weighting in
several places and as a result obtain different estimates of $\fnlloc$.
Furthermore, in both \cite{Komatsu:2008hk,Yadav:2007yy} the uncertainty $\sigma(\fnlloc)$ decreases with $\ellmax$ for $\ellmax\lesssim 400$ and then slightly 
increases for larger $\ellmax$, making it unclear what is the best-motivated choice of $\ellmax$ for a ``bottom-line'' estimate of $\fnlloc$.
In constrast, the optimal estimator is unique and $\sigma(\fnlloc)$ is a decreasing function of $\ellmax$ which eventually saturates.

\subsection{Foreground marginalization}
\label{ssec:foreground_marginalization}

We use the WMAP foreground model: the total foreground contribution to the temperature at frequency $\nu$ in pixel $\nu$ is given by
\be
T_{\rm fg}(\n) = b_1(\nu) T_{\rm synch}(\n) + b_2(\nu) T_{\rm ff}(\n) + b_3(\nu) T_{\rm dust}(\n)
\ee
where the functions $b_i(\nu)$ encode the frequency dependence of the foregrounds, and
$T_{\rm synch}(\n)$, $T_{\rm ff}(\n)$, $T_{\rm dust}(\n)$ are spatial templates for synchrotron, free-free and dust emission.
For more details, including construction of the spatial templates and the procedure for estimating $b_i(\nu)$, see \cite{Gold:2008kp}.

The WMAP data release includes ``clean maps'' which are obtained by subtracting $T_{\rm fg}(\n)$ from the ``raw maps'' which are directly observed in each channel.
In \cite{Komatsu:2008hk}, the $\fnlloc$ estimator was applied to clean maps, assuming that any systematic error from foregrounds is small.
(This assumption is tested by checking frequency dependence and dependence on the mask.)
In \cite{Yadav:2007yy}, foregrounds were treated by applying the estimator to the raw maps, and assuming that any bias due
to foregrounds is negative.
Under this assumption, the raw-map estimate is a lower bound on $\fnlloc$ even in the presence of foregrounds.

Using the optimal estimator, there is a third possibility for handling foregrounds: one can marginalize over the templates by modifying
the noise covariance $N$ so that each template $T_i(\n)$ is assigned infinite variance.
The optimal estimator in Eq.~(\ref{eq:Edef}) then estimates $\fnlloc$ in a way which is ``blind'' to the amplitude of the template modes in the data.
The variance of the estimator will be slightly increased to account for this loss of information.
We marginalize the templates independently in each WMAP channel to avoid making any assumptions about the functions $b_i(\nu)$.
Using the foreground-marginalized optimal estimator, direct template cleaning of the maps is not necessary: the $\fnlloc$ estimates from raw
and clean maps will be the same.
This estimator will be our default choice in the rest of the paper unless otherwise specified.

\section{Results}\label{results}

\begin{figure}
\centerline{\epsfxsize=12cm\epsffile{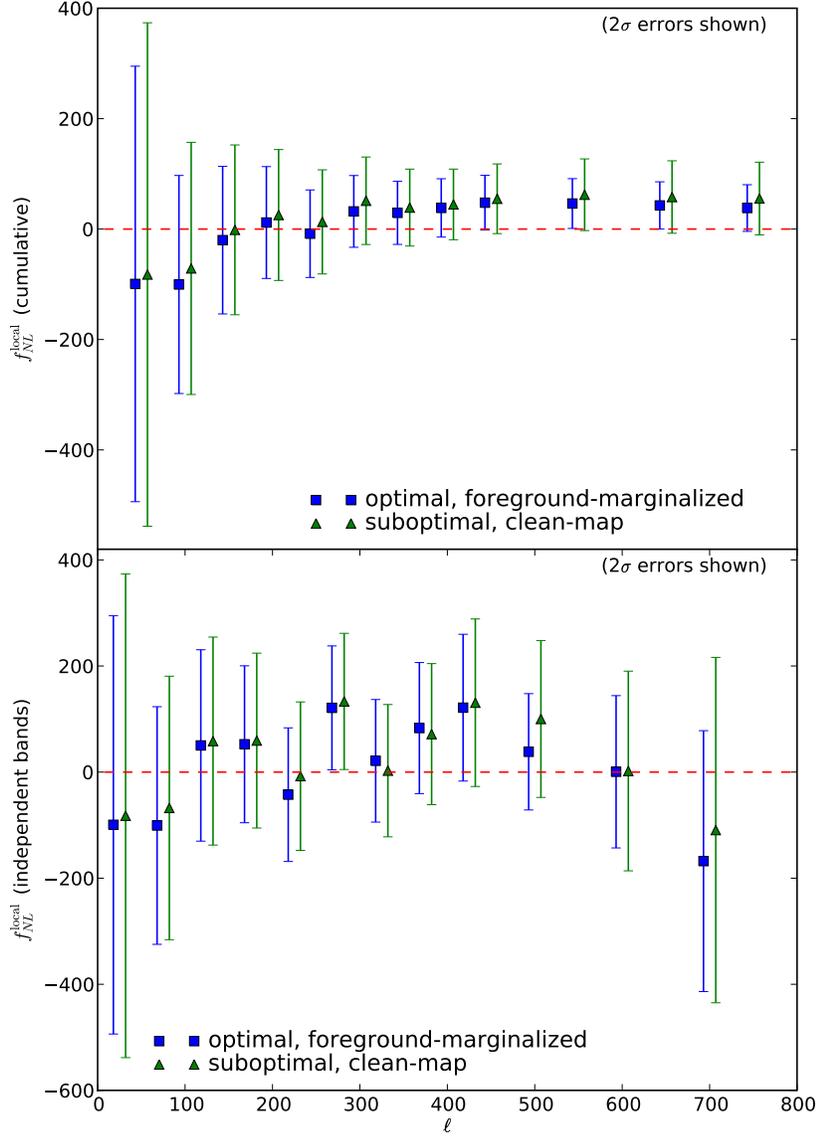}}
\caption{Constraints on $\fnlloc$ using 5-year data and KQ75 mask, 
using both the optimal estimator (squares) and the old estimator applied to clean maps (triangles).
The top panel shows cumulative results (constraints using all the information up to a given $\ell$) 
while the bottom one shows contributions from separate $\ell$ bins.
Our overall $\fnlloc$ estimate, taking $\ellmax=750$, is $(38\pm 21)$ for the optimal estimator and $(55\pm 33)$ for suboptimal.}
\label{oldvsoptimal}
\end{figure}

Our best constraint on $\fnlloc$ comes from our optimal analysis applied to WMAP 5 year data using the foreground marginalization technique. 
We find: $\fnlloc=(38 \pm 21)$ at 1$\sigma$.
The primordial fluctuations are consistent with Gaussian,  $-4< \fnlloc < 80$ at 95\% CL.
It is important to point out that our analysis of the WMAP data results in error bars that are smaller than previous analyses as a result of our using the optimal estimator. 
This can be clearly seen in Figure~\ref{oldvsoptimal} where we directly compare optimal and suboptimal estimators for otherwise the same choices of analysis parameters. 
Note that the optimal estimator is already better at even relatively large scales, but it gets substantially better for large values of $\ellmax$.

Given the importance of a detection of non-Gaussianity it is important to understand how robust our results are to various choices of analysis parameters and data. We explore these issues in the next two sections.

\section{WMAP 5 vs WMAP 3}\label{tests1}

The differences between the results of \cite{Komatsu:2008hk} and \cite{Yadav:2007yy} might lead to the suspicion that something changed in the 
data between the 5 and 3 year data release. 

When we analyze the 3-year dataset with the optimal foreground-marginalized estimator, we find $\fnlloc=(58\pm 23)$.
Thus, between the 3-year and 5-year datasets, we find a shift $\Delta\fnlloc = -20$ in the value of $\fnlloc$.
This may seem too large to be a statistical event, given that the error on $\fnlloc$ is not much better in the 5-year data than in the 3-year data.
What is the cause of this shift, and is it consistent with statistics?

\begin{figure}[!ht]
\centerline{\epsfxsize=12cm\epsffile{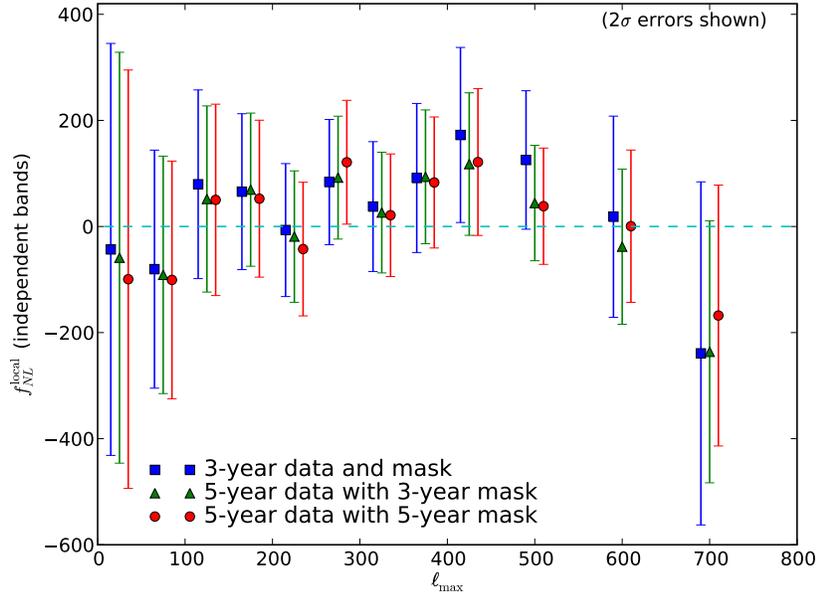}}
\caption{Estimates of $\fnlloc$ using the optimal foreground-marginalized estimator with
3-year data and mask (squares), 5-year data with 3-year mask (triangles), and 5-year data and mask (circles).
We have shown the contributions from separate $\ell$ bins; the overall estimates of $\fnlloc$ obtained by summing all bins
are $(58\pm 23)$, $(37\pm 21)$ and $(38\pm 21)$ respectively.}
\label{5vs3}
\end{figure}

There are four differences between the 3-year and 5-year datasets which are relevant for our $\fnlloc$ analysis:
different maps (3 years of data vs 5), different foreground mask (Kp0 vs KQ75), different beams, and different best-fit cosmological parameters.
We find that the changes to the beams and cosmological parameters have a negligible effect on $\fnlloc$.
In Figure~\ref{5vs3}, we compare the changes due to the updated maps and updated mask.
Splitting the overall $\fnlloc$ estimate into independent $\ell$ bins, it is seen that the change to $\fnlloc$ is mainly coming
from $\ell\approx 450$ where it is mostly due to the updated maps.

As a test for systematics, we generated ``paired'' simulations of the 3-year and 5-year datasets.
Each pair consists of a 3-year simulation and a 5-year simulation which share the same CMB realization.
The noise realization in each 5-year simulation is constructed by combining the noise realization from the corresponding 3-year
simulation with an independent noise realization corresponding to 2 years of integration time, in a way which mimics the way the
data from different years is combined in the real WMAP data.
We find that the RMS $\Delta\fnlloc$ between the 3-year simulations and the 5-year simulations is 16, so the shift observed in
the data is within statistics.
The same is true for each individual $\ell$ bin in Figure~\ref{5vs3}.
We conclude that there is nothing dramatically different in the two data sets.

This comparison between WMAP3 and WMAP5 assumes the optimal estimator.
If the suboptimal estimator is used instead, we also find an RMS $\Delta\fnlloc$ between 3-year and 5-year which is equal to 16,
so the difference between the WMAP3 result reported in \cite{Yadav:2007yy} and the WMAP5 result reported in \cite{Komatsu:2008hk}
($\fnlloc = 87 \pm 30$ and $\fnlloc = 58\pm 36$ respectively, for large $\ellmax$) is marginally consistent with being a statistical event.

One puzzling feature of the 3-year dataset is the large value of $\fnlloc$ reported in \cite{Yadav:2007yy} at $\ellmax=750$, compared to estimates at
$\ellmax=350$ which had been reported previously \cite{Creminelli:2006rz,Spergel:2006hy}.
Using our pipeline, we see a change of $\Delta\fnlloc=20$ (optimal estimator) or $\Delta\fnlloc=32$ (suboptimal estimator) between these values of
$\ellmax$, a less dramatic shift than the $\Delta\fnlloc=52$ change reported in \cite{Yadav:2007yy}.
In simulation we find that the RMS change in $\fnlloc$ between these values of $\ellmax$ is
$\approx 19$, so it appears difficult to interpret a change as large as 52 as a statistical event.
(We find an RMS change $\approx$19 for both the optimal or suboptimal estimator, and with either 3-year or 5-year data, so it seems to
be a robust quantity.)

\begin{figure}
\centerline{\epsfxsize=12cm\epsffile{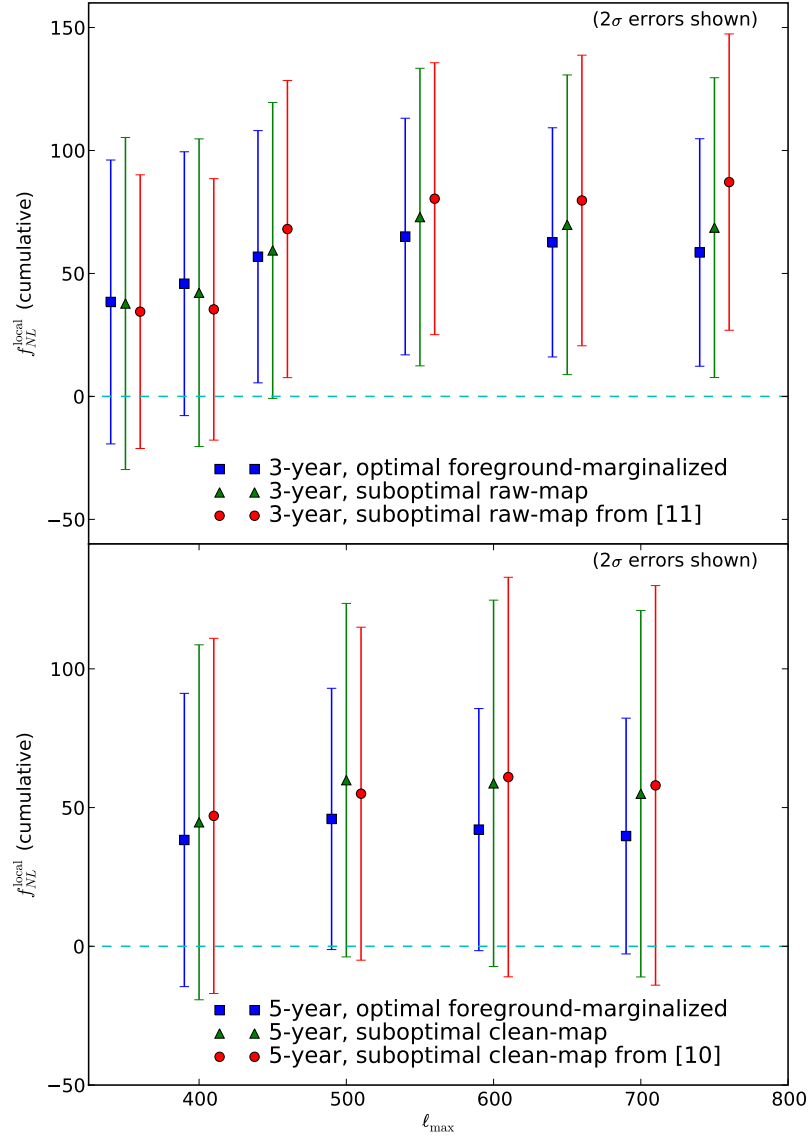}}
\caption{{\em Top panel:} 
Comparison between 3-year results reported in \cite{Yadav:2007yy}
and results obtained from our pipeline, using either the optimal or suboptimal estimator.
We apply the suboptimal estimator to 3-year raw maps for consistency with \cite{Yadav:2007yy}.
{\em Bottom panel:} Comparison between 5-year results (optimal estimator, raw maps)
reported in \cite{Komatsu:2008hk} and results obtained from our pipeline using the
optimal or suboptimal estimator.
We apply the suboptimal estimator to 5-year clean maps for consistency with \cite{Komatsu:2008hk}.}
\label{fig:comparison}
\end{figure}

There also appears to be some systematic tendency for our pipeline to produce lower $\fnlloc$ values, compared to the 3-year results of \cite{Yadav:2007yy},
at large $\ellmax$ (Figure~\ref{fig:comparison}, top panel).
If we take $\ellmax=750$ and apply the suboptimal estimator to 3-year raw maps for comparison with \cite{Yadav:2007yy}, we get $\fnlloc=69\pm 30$ 
(disfavoring $\fnlloc=0$ at 2.3$\sigma$), whereas $\fnlloc=87\pm 30$ (disfavoring $\fnlloc=0$ at 2.9$\sigma$) was reported in \cite{Yadav:2007yy}.
The reason for this disagreeemnt is unclear, but may simply be the result of making different choices of weighting in the suboptimal estimator.
This is good motivation for using the optimal estimator, which is unique and therefore different implementations should agree precisely.

The agreement between our pipeline and the 5-year results from \cite{Komatsu:2008hk} is better (Figure~\ref{fig:comparison}, bottom panel). Furthermore we have internally compared the non-optimal pipeline in this paper with the one we used in \cite{Creminelli:2006rz} and found  them to agree at the percent level. Both pipelines were independently developed. 

\section{Foregrounds}
\label{tests2}

\subsection{Large-scale galactic foregrounds}
\label{ssec:large_scale_fg}

\begin{figure}[!ht]
\centerline{\epsfxsize=12cm\epsffile{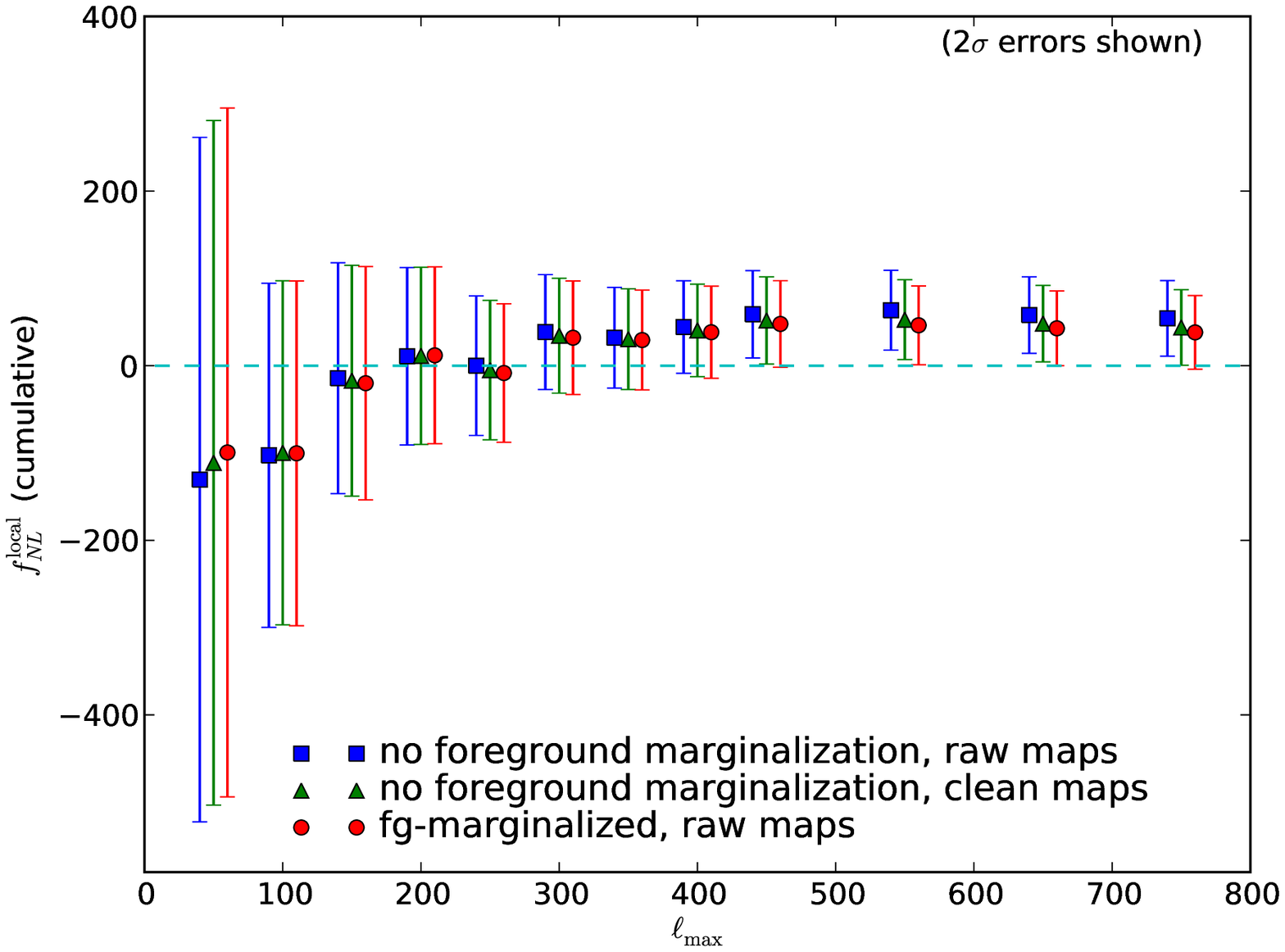}}
\caption{Comparison between the foreground-marginalized optimal estimator ($\hfnl$), and the optimal estimator without foreground marginalization
($\hfnl^0$) applied to either raw or clean 5-year maps with KQ75 mask.}
\label{fig:raw_vs_clean}
\end{figure}

Perhaps the most worrying systematic effect in this analysis is non-Gaussian contamination by foregrounds.
Let us denote the foreground-marginalized optimal estimator by $\hfnl$, and the optimal estimator constructed 
without marginalizing foregrounds by $\hfnl^0$.
We can get a crude idea of how important foregrounds are, at the order-of-magnitude level, by comparing the 
raw-map value of $\hfnl^0$, the clean-map value of $\hfnl^0$, and the value of $\hfnl$.  (As described in 
\S\ref{ssec:foreground_marginalization}, $\hfnl$ gives the same value when applied to raw or clean maps.)
In the five-year dataset with KQ75 mask, we find that $\hfnl$ agrees well with $\hfnl^0({\rm clean})$, and $\hfnl^0({\rm raw})$ is larger
by $\approx$10 (Figure~\ref{fig:raw_vs_clean}).

This result suggests that foreground contamination is mild if the KQ75 mask is used, but also raises a puzzle.
In \cite{Yadav:2007yy} it was argued that foregrounds always make a negative contribution to $\fnlloc$ even in a single
realization, so that the raw-map estimate can be taken as a lower bound on the true value of $\fnlloc$.
(For example, the systematic error due to foregrounds outside the Kp0 mask is quoted as a ``one-sided'' range ${}_{-0}^{+6}$.)
However, with the optimal estimator and 5-year dataset, we find that $(\hfnl^0({\rm raw})-\hfnl^0({\rm clean}))$
is positive.  Is this a sign that something is wrong with the estimator, or the foreground model?
What general statement can we make about the sign of $(\hfnl^0({\rm raw})-\hfnl^0({\rm clean}))$?

The raw maps, clean maps, and foreground maps are related (with $C^{-1}$ filter applied) by:
\be
(C^{-1}a)_{\ell m}^{\rm raw} = (C^{-1}a)_{\ell m}^{\rm clean} + (C^{-1}a)^{\rm fg}_{\ell m}
\ee
Using Eq.~(\ref{eq:Edef}), we can write $(\hfnl^0({\rm raw})-\hfnl^0({\rm clean}))$ as the sum of three terms:
\be
\hfnl^0({\rm raw}) - \hfnl^0({\rm clean}) = (FTT) + (FFT) + (FFF)
\ee
where we have defined:
\ba
FTT &=& \frac{3}{\mathcal N} B_{\ell_1\ell_2\ell_3} \threej{\ell_1}{\ell_2}{\ell_3}{m_1}{m_2}{m_3}
       \Big[ (C^{-1}a)_{\ell_1 m_1}^{fg} (C^{-1}a)_{\ell_2 m_2}^{\rm clean} (C^{-1}a)_{\ell_3 m_3}^{\rm clean}
         - C^{-1}_{\ell_1 m_1, \ell_2 m_2} (C^{-1}a)_{\ell_3 m_3}^{\rm fg} \Big]   \nn \\
FFT &=& \frac{3}{\mathcal N} B_{\ell_1\ell_2\ell_3} \threej{\ell_1}{\ell_2}{\ell_3}{m_1}{m_2}{m_3}
       (C^{-1}a)_{\ell_1 m_1}^{fg} (C^{-1}a)_{\ell_2 m_2}^{\rm fg} (C^{-1}a)_{\ell_3 m_3}^{\rm clean}  \nn \\
FFF &=& \frac{1}{\mathcal N} B_{\ell_1\ell_2\ell_3} \threej{\ell_1}{\ell_2}{\ell_3}{m_1}{m_2}{m_3}
       (C^{-1}a)_{\ell_1 m_1}^{fg} (C^{-1}a)_{\ell_2 m_2}^{\rm fg} (C^{-1}a)_{\ell_3 m_3}^{\rm fg}  \label{eq:fff}
\ea
(Note that we have chosen to include the linear term in the FTT piece.  This is the most natural choice since
it ensures that $\langle FTT \rangle = 0$, where the expectation value is taken over random CMB realizations
with the foreground template fixed.)

For the WMAP foreground model and five-year dataset, we find (FTT)=10.4, (FFT)=-0.1, and (FFF)=-0.2:
the shift in $\fnlloc$ between the raw and clean maps is entirely due to the (FTT) term.
This term represents accidental correlation between large-scale foregrounds and small-scale non-foreground power,
and is equally likely to be positive or negative.
We conclude that it is not safe to assume that $(\hfnl^0({\rm raw})-\hfnl^0({\rm clean}))$ is negative.
The value depends on the way that the small-scale modes are filtered and can be different for the
optimal and suboptimal estimators.

In this paper, we will treat large-scale foregrounds by marginalizing the template amplitudes in the
optimal estimator as described in \S\ref{ssec:foreground_marginalization}.
When we use the suboptimal estimator, we will simply analyze clean maps and neglect the (small) extra 
uncertainty in $\fnlloc$ due to uncertainty in the template amplitudes.
In principle, one could estimate $\fnlloc$ from raw maps and treat the zero-mean (FTT) term as
systematic error from foregrounds, by increasing the uncertainty $\sigma(\fnlloc)$.
However, assuming the WMAP foreground model, we note that the size of the (FTT) term in the WMAP data is 10.4,
whereas the RMS value in simulation is 4.0.
(The simulations were constructed by evaluating $(FTT)$ in Eq.~(\ref{eq:fff}) using a random CMB realization
and the WMAP foreground templates.)
The larger value seen in the data can be interpreted as a test for foreground contamination that is
failing at 2.5$\sigma$, but it is unclear how to interpret this further.

\subsection{Small-scale galactic foregrounds}

The WMAP foreground templates are smoothed with a $1^\circ$ beam and therefore
the template-marginalization procedure does not remove foregrounds on small angular scales. 
This may contaminate the $\fnlloc$ estimator, which is a cross-correlation between
long-wavelength temperature $(\ell\approx 20)$ and small-scale power $(\ell\approx 350)$.
Such contamination, if present, is not included in the analysis from the preceding subsection, in which it is assumed
that the templates agree perfectly with the real foregrounds.

We can roughly estimate the contamination due to uncleaned small-scale galactic foregrounds in the following way.
The dust template is available at high resolution \cite{Schlegel:1997yv}, so we can define a ``clean+'' map for each WMAP channel by:
\be
T_{\rm clean+}(\n) = T_{\rm raw}(\n) - b_1(\nu) T_{\rm synch}^{\rm smth}(\n) - b_2(\nu) T_{\rm ff}^{\rm smth}(\n) - b_3(\nu) T_{\rm dust}^{\rm hires}(\n)
\ee
where ``smth'' denotes a foreground template  smoothed with a $1^\circ$ beam, and ``hires'' denotes a template smoothed with the instrumental beam.
The clean and clean+ maps agree on large scales, but on small scales, foregrounds have been partially subtracted in the clean+ maps.
We find that the $\fnlloc$ estimates from clean and clean+ maps are negligibly different ($\Delta\fnlloc \lesssim 0.75$ in all $\ell$ bands).
Our clean+ maps only include small-scale foreground contributions from dust;
however, dust is expected to be the largest small-scale foreground in W-band, and comparable to the other foregrounds in V-band.
(We checked this using the MEM maps from the WMAP 5-year release.)
This suggests that the impact of small-scale galactic foregrounds is negligible.

\subsection{Point sources}

The largest source of small-scale foreground power in WMAP is from unresolved point sources, mainly radio sources.
If point sources are assumed isotropic and unclustered, it has been shown \cite{Komatsu:2003iq} that the $\fnlloc$ contamination is negligible.
However, the clustering of sources in WMAP is not well characterized, and the number density of unresolved sources may be larger
near the galactic plane, and is certainly larger in the ecliptic plane, where the larger noise level in WMAP makes it harder to 
detect and mask sources.
For this reason, we would like to do a direct test for point source contamination by deliberately weakening the point source mask
and comparing the value of $\fnlloc$.

A second concern, pointed out in \cite{Komatsu:2008hk}, is that the WMAP source detection procedure has a lower effective flux threshhold in
regions where the local CMB temperature is higher than average.  This may negatively correlate the level of unresolved point source
power to the CMB temperature and fake a positive-$\fnlloc$ signal.  To address this concern at the same time, we construct a ``KQ75-CW''
mask by leaving the galactic part of the KQ75 mask unchanged, but replacing the source mask by one constructed from the Chen \& Wright
\cite{Chen:2008gw} catalog, in which only difference maps between WMAP frequences are used to detect sources.
Our KQ75-CW mask contains $\approx$40\% as many sources as the KQ75 mask (this is largely due to the use of sources from external
catalogs in KQ75 \cite{Bennett:2003ca}), but the unresolved sources should be uncorrelated to the CMB.

Applying the optimal foreground-marginalized estimator to the five-year data using both the KQ75 and KQ75-CW masks, we find that
the difference in $\fnlloc$ is very small ($\Delta\fnlloc \lesssim 2$; the precise value depends on $\ellmax$).
We conclude that any bias due to correlations between the KQ75 point source mask and the CMB is negligible.
This test also suggests (but does not prove) that any bias due to unresolved point sources is small, since most of the sources 
masked by KQ75 are not masked by KQ75-CW.
The systematic error from unresolved sources was calculated by the WMAP team in \cite{Komatsu:2008hk} using Monte Carlo simulations,
and found to be small compared to the statistical error.

\subsection{Other tests for foreground contamination}

\begin{figure}[!ht]
\centerline{\epsfxsize=12cm\epsffile{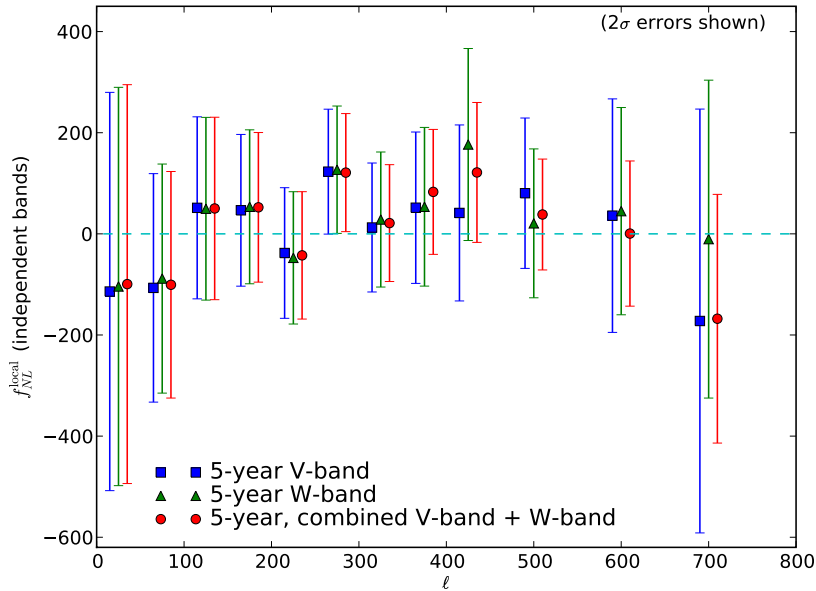}}
\caption{Comparison between V-band data, W-band data, and (V+W) combined, using the optimal foreground-marginalized estimator, five-year data, and KQ75 mask.
We show the contributions to $\fnlloc$ from independent $\ell$ bins.}
\label{fig:frequency_dependence}
\end{figure}

In Figure~\ref{fig:frequency_dependence}, we compare the optimal foreground-marginalized estimator using five-year V-band data, W-band data,
and the combined (V+W) result shown previously.  The three cases agree well at low $\ell$ where the data is signal-dominated, and
deviate somewhat at high $\ell$ where the noise realizations in V-band and W-band are independent.  The differences between V-band
and W-band are consistent with simulation, i.e. no evidence is seen for a frequency-dependent signal.

\begin{figure}[!ht]
\centerline{\epsfxsize=12cm\epsffile{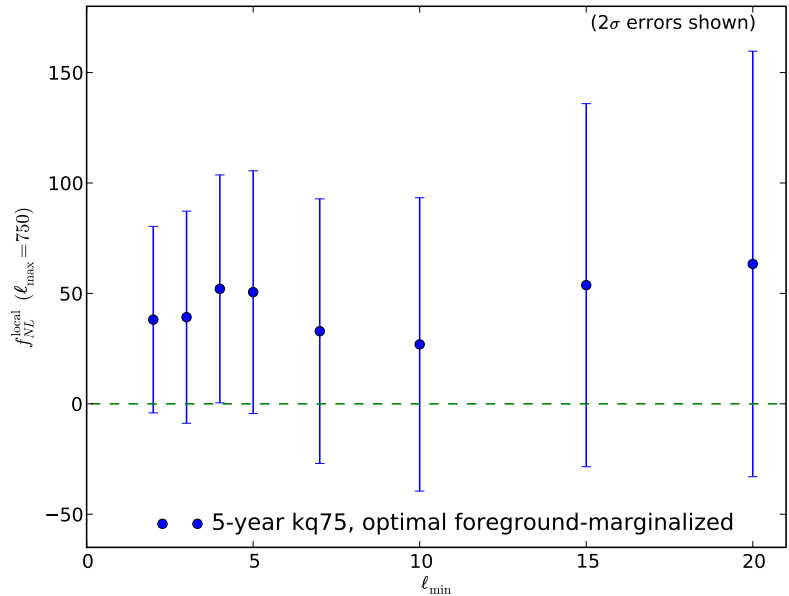}}
\caption{$\fnlloc$ estimates with varying minimum multipole $\ellmin$, using the optimal foreground-marginalized estimator, five-year data, KQ75 mask,
and taking $\ellmax=750$ throughout.}
\label{fig:varying_lmin}
\end{figure}

Because foregrounds in WMAP are most important on the largest scales (particularly $\ell=2$),
another test we can do for foreground contamination is to fix $\ellmax=750$ and vary the minimum multipole $\ellmin$ that is used to estimate $\fnlloc$.
One of the most striking results reported in \cite{Yadav:2007yy} is that even with $\ellmin=20$, where about half the statistical weight is lost,
evidence for positive $\fnlloc$ is still seen in three-year raw maps at high significance: $\fnlloc=135\pm 48$ at 1$\sigma$.
We do not see such a signal, finding $\fnlloc=48\pm 56$ (suboptimal estimator, raw maps) or $\fnlloc=50\pm 51$ (optimal estimator, foreground-marginalized)
using three-year data, Kp0 mask, $\ellmin=20$ and $\ellmax=750$.
Results for five-year data are shown in Figure~\ref{fig:varying_lmin}.  The statistical significance of nonzero $\fnlloc$ stays roughly constant
out to $\ellmin \approx 6$, and then decreases.

\section{Summary}\label{conclusions}

We have applied the optimal estimator of \cite{Creminelli:2005hu}, with the implementation developed in \cite{Smith:2006ud}, to the 5 year WMAP data. 
Our results are summarized in Table~\ref{tab:summary}.
Marginalizing over the amplitude of foreground templates we get 
$-4< \fnlloc < 80$ at 95\% CL. Error bars of previous analysis are roughly 40\% larger than these. The probability that a Gaussian simulation, analyzed using our estimator, gives a result larger in magnitude than the one we find is 7\,\%. 

We did extensive tests of our results including implementing our own sub-optimal estimator to compare with published results. We concluded that:
\begin{itemize}
\item The optimal estimator outperforms the sub-optimal one both on large and small angular scales.  
\item The differences we see between the results of our pipeline when applied to  the three and five year WMAP data releases are consistent with being statistical fluctuations. 
\item Our implementation of the sub-optimal estimator is in good agreement with the results obtained by the WMAP team in \cite{Komatsu:2008hk}. We see see some small but significant discrepancies with the results of Yadav \& Wandelt \cite{Yadav:2007yy} on small angular scales. We see no significant differences in excess of noise between the results of our optimal and sub-optimal estimators.  
\item The fluctuations we see between results at different $\ell$ bands are consistent with being statistical.  
\item After foreground template marginalization we do not see any evidence of foregrounds affecting our results in a significant way. We see no evidence for frequency dependence in our results and our constraints are robust to the choice of minimum $\ell$ value allowed in the triangles. 
\item The foreground contamination present outside of KQ75 in the raw maps seem to create a fake non-Gaussian signal that shifts the mean value of $\fnlloc$ up  by about half a $\sigma$.  This happens through an accidental correlation between the CMB and foreground signals. This term does not have a definite sign, it is realization dependent.  We conclude that using raw maps instead of foreground marginalized maps does not generally guarantee getting a conservative lower limit on $\fnlloc$. 
\item The lack of small scale power in the foreground templates does not appear to bias our results in any noticeable way. Our results are robust to the choice of point source mask.  
\end{itemize}

\begin{table}
\begin{center}
\begin{tabular}{|cc|c|c|c|c|c|c|c|}
\hline\hline \multicolumn{8}{|c|}{Optimal estimator} \\ \hline\hline
\multicolumn{2}{|c|}{Dataset + mask + fg} & $\ell_{\rm max}=250$ & 350 & 450 & 550 & 650 & 750 \\ \hline
5-yr V+W & KQ75 fg-marg. & $-8 \pm 40$ & $29\pm 29$ & $48 \pm 25$ & $46 \pm 23$ & $43 \pm 22$ & $38 \pm 21$ \\
5-yr V+W & KQ75 clean-map & $-5\pm 40$ & $30 \pm 29$ & $52 \pm 25$ & $53 \pm 23$ & $48 \pm 22$ & $44 \pm 21$ \\
5-yr V+W & KQ75 raw-map & $0 \pm 40$ & $32 \pm 29$ & $59 \pm 25$ & $63 \pm 23$ & $57 \pm 22$ & $54 \pm 21$ \\
3-yr V+W & kp0 fg-marg. & $19\pm 40$ & $38\pm 29$ & $57\pm 26$ & $65\pm 24$ & $63\pm 23$ & $59\pm 23$ \\
5-yr V+W & kp0 fg-marg. & $8\pm 39$ & $33\pm 28$ & $51\pm 24$ & $50\pm 22$ & $44\pm 21$ & $38\pm 21$ \\
5-yr V & KQ75 fg-marg. & $-9 \pm 40$ & $26 \pm 30$ & $31 \pm 27$ & $36 \pm 25$ & $36 \pm 25$ & $34 \pm 25$ \\
5-yr W & KQ75 fg-marg. & $-8 \pm 41$ & $31 \pm 30$ & $44 \pm 28$ & $41 \pm 26$ & $41 \pm 26$ & $41 \pm 25$ \\ \hline\hline
\multicolumn{8}{|c|}{Suboptimal estimator (same parameters as \cite{Yadav:2007yy})} \\ \hline\hline
\multicolumn{2}{|c|}{ } & $\ell_{\rm max}=250$ & 350 & 450 & 550 & 650 & 750 \\ \hline
3-yr V+W & kp0 raw-map & $31 \pm 45$ & $38\pm 34$ & $59\pm 30$ & $73\pm 30$ & $70\pm 30$ & $69\pm 30$ \\ \hline\hline
\multicolumn{8}{|c|}{Suboptimal estimator (same parameters as \cite{Komatsu:2008hk})} \\ \hline\hline
\multicolumn{2}{|c|}{ } & $\ell_{\rm max}=200$ & 300 & 400 & 500 & 600 & 700 \\ \hline
5-yr V+W & kq75 clean-map & $25\pm 59$ & $51\pm 40$ & $45\pm 32$ & $60\pm 32$ & $59\pm 33$ & $55\pm 33$  \\ \hline\hline
\end{tabular}
\end{center}
\caption{Estimated values of $\fnlloc$, with $1\sigma$ errors, for various choices of dataset, mask, and foreground cleaning procedure used throughout this paper.}
\label{tab:summary}
\end{table}

\section*{Acknowledgments}
We would like to thank Eiichiro Komatsu and David Spergel for help during the project.  
KMS was supported by an STFC Postdoctoral Fellowship.
LS was supported in part by the National Science Foundation under Grant No. PHY-0503584. MZ was supported by NASA NNG05GJ40G and NSF AST-0506556 as well as the David and Lucile Packard, Alfred P. Sloan and John D. and Catherine T. MacArthur foundations. 
KMS would like to thank the hospitality of the Department of Astrophysics at Princeton University, where this work was partially carried out.

\appendix

\section{Implementation of the optimal estimator}
\label{app:estimator}

In this appendix, we describe some implementational details of our analysis pipeline.

\subsection{$C^{-1}$ implementation}

We represent the WMAP data as a length-$\Npix$ data vector $d_i$ and an $(\Npix)$-by-$(\Npix)$ noise 
covariance matrix $N_i$, for each WMAP channel $i=1,2,\ldots,\Nchan$.
In the WMAP noise model, different pixels are uncorrelated, i.e. each $N_i$ is a diagonal matrix.
We represent the CMB realization in harmonic space by a length-$\Nalm$ vector $a$, where $\Nalm$ is the
number of linearly independent multipoles $a_{\ell m}$ such that $\ell\le\ellmax = 1000$.
For each channel, we introduce an $\Npix$-by-$\Nalm$ matrix $A_i$ which combines the beam convolution
and spherical transform operations, so that we can write:
\be
d_i = A_i a + n_i   \label{eq:cinv_setup}
\ee
where the noise $n_i$ is a length-$\Npix$ vector.
We write the signal covariance corresponding to the fiducial power spectrum as an $\Nalm$-by-$\Nalm$
matrix $S$ (thus $\langle aa^T \rangle = S$ and $\langle n_i n_j^T\rangle = N_i \delta_{ij}$).

Given the data in Eq.~(\ref{eq:cinv_setup}), there is an optimal (minimum-variance) map $\ha$ \cite{Tegmark:1996qs},
which is an unbiased estimator of the CMB realization $a$:
\ba
\ha &=& a + \eta  \\
\langle \eta \eta^T \rangle &=& N
\ea
The optimal map $\ha$ and its $\Nalm$-by-$\Nalm$ noise covariance $N$ are jointly defined by:
\ba
N^{-1} \ha &=& \sum_{i=1}^{\Nchan} A_i^T N_i^{-1} d_i  \label{eq:cinv_ninva} \\
N^{-1} &=& \sum_{i=1}^{\Nchan} A_i^T N_i^{-1} A_i  \label{eq:cinv_ninv}
\ea
Going from the data in Eq.~(\ref{eq:cinv_setup}) to the optimal map $\ha$ does not lose information, so one can think of the data
as if the CMB realization $a$ were directly observed in harmonic space, with noise covariance $N$ defined by Eq.~(\ref{eq:cinv_ninv}).

To evaluate the optimal $\fnlloc$ estimator (Eq.~(\ref{eq:Edef})), we need to compute $(S+N)^{-1}\ha$.
Formally, this is straightforward using Eqs.~(\ref{eq:cinv_ninva}),~(\ref{eq:cinv_ninv}), but in practice there are two computational obstacles:
\begin{enumerate}
\item The WMAP resolution is too large for dense $(\Npix)$-by-$(\Npix)$ linear algebra (or $(\Nalm)$-by-$(\Nalm)$ linear algebra)
to be computationally feasible.
\item The inverse noise covariance $N^{-1}$ will usually not be invertible.
\end{enumerate}
To explain the second problem better, we note that in our pipeline, we represent the sky cut by assigning infinite noise to the pixels which are masked.
Therefore the inverse noise covariance $N_i^{-1}$ in pixel space is a non-invertible diagonal matrix (the entries corresponding
to masked pixels are zero).
Going to harmonic space, the operator $N^{-1}$ and the vector $N^{-1}\ha$ in Eqs.~(\ref{eq:cinv_ninva}),~(\ref{eq:cinv_ninv})
are still defined, but $N$ and $\ha$ generally will not be, because $N^{-1}$ is not invertible.

To solve the first problem, we first observe that there is a computationally efficient procedure for multiplying a length-$\Nalm$ 
vector by the matrix $N^{-1}$.
This follows from Eq.~(\ref{eq:cinv_ninv}), since multiplication by the operator $A_i$ (or $A_i^T$) can be done using a fast spherical transform, and
multiplication by $N_i^{-1}$ is trivial because $N_i$ is a diagonal matrix in the WMAP noise model.
The next step is to write
\be
(S+N)^{-1} \ha = S^{-1} (S^{-1} + N^{-1})^{-1} N^{-1} \ha    \label{eq:cinv_cg}
\ee
Since we can compute $N^{-1}\ha$ using Eq.~(\ref{eq:cinv_ninva}), the only missing ingredient is an efficient procedure for multiplying
a vector by the matrix $(S^{-1}+N^{-1})^{-1}$.
Note that we have already described an efficient procedure for performing the ``forward'' operation $(S^{-1}+N^{-1})$.
Given such a procedure, conjugate gradient inversion \cite{NR} is a well-known iterative 
method for performing the inverse operation $(S^{-1}+N^{-1})^{-1}$
which avoids direct matrix inversion.
Obtaining rapid convergence with conjugate gradient inversion usually depends on constructing a good preconditioner, or approximate inexpensive inverse operation.
Using the multigrid preconditioner from \cite{Smith:2007rg}, the computational cost of each $C^{-1}$ operation is about 10 CPU-minutes for the WMAP5 V+W dataset.

This solution to the first problem above (infeasibility of dense linear algebra) also solves the second problem (noninvertibility of $N^{-1}$), 
since evaluating the right-hand side of Eq.~(\ref{eq:cinv_cg}) by conjugate gradient inversion only requires us to compute
$N^{-1}\ha$, and to multiply vectors by $N^{-1}$.
In fact, we have deliberately written Eq.~(\ref{eq:cinv_cg}) in such a way that $\{N,\ha\}$ have been eliminated in favor of $\{N^{-1},N^{-1}\ha\}$.

Finally, we describe our implementation of template marginalization.
We always marginalize templates in pixel space, and do the marginalization independently for each WMAP channel.
Some results in this paper include foreground marginalization, i.e. we have marginalized three modes corresponding to the synchrotron, free-free 
and dust foregrounds.
In addition, all results which use the optimal estimator include marginalization of the four modes corresponding to the monopole and dipole.
Formally, let $\tau$ be an $N_{\rm tmpl}$-by-$\Npix$ matrix which contains the pixel-space templates.
If we denote the pixel-space inverse noise covariance with and without template marginalization by $N_i^{-1}$ and ${\bar N_i}^{-1}$ respectively,
then the two are related by:
\be
N_i^{-1} = \lim_{\eta\rightarrow\infty} [ \bar N_i + \eta \tau^T \tau ]^{-1} 
  = {\bar N_i}^{-1} - {\bar N_i}^{-1} \tau^T [\tau {\bar N_i}^{-1} \tau^T]^{-1} \tau {\bar N_i}^{-1}  \label{eq:marg}
\ee
The method described above for computing $C^{-1}\ha$ only requires us to have a procedure for multiplying a vector by $N_i^{-1}$.
This is straightforward using the right-hand side of Eq.~(\ref{eq:marg}), since $\bar N_i$ is diagonal and the rest of the matrices are small enough that
dense linear algebra is computationally feasible.

\subsection{$\fnlloc$ estimator}

The preceding subsection describes our algorithm for computing $C^{-1}\ha$, where $\ha_{\ell m}$ is a minimum-variance map
made by optimally combining all WMAP channels.  The optimal $\fnlloc$ estimator is obtained from this as follows:
\be
\hfnl = \frac{1}{\mathcal N} B_{\ell_1\ell_2\ell_3} \threej{\ell_1}{\ell_2}{\ell_3}{m_1}{m_2}{m_3}
       \Big[ (C^{-1}\ha)_{\ell_1 m_1} (C^{-1}\ha)_{\ell_2 m_2} (C^{-1}\ha)_{\ell_3 m_3} 
         - 3 C^{-1}_{\ell_1 m_1, \ell_2 m_2} (C^{-1}\ha)_{\ell_3 m_3} \Big]  \label{eq:Edef2}
\ee
where $B_{\ell_1\ell_2\ell_3}$ is the local bispectrum normalized to $\fnlloc=1$.

We have written the estimator in harmonic space where it is simplest, but in practice the cubic term
is evaluated efficiently in position space using the KSW construction \cite{Komatsu:2003iq}.  The linear term is obtained
as a Monte Carlo average following \cite{Creminelli:2005hu}:
\ba
\hfnl &=& \frac{1}{\mathcal N} B_{\ell_1\ell_2\ell_3} \threej{\ell_1}{\ell_2}{\ell_3}{m_1}{m_2}{m_3}
            (C^{-1}\ha)_{\ell_1 m_1} (C^{-1}\ha)_{\ell_2 m_2} (C^{-1}\ha)_{\ell_3 m_3}  \label{eq:Edef3} \\
&& - \left\langle \frac{3}{\mathcal N} B_{\ell_1\ell_2\ell_3} \threej{\ell_1}{\ell_2}{\ell_3}{m_1}{m_2}{m_3}
         (C^{-1}\hs)_{\ell_1 m_1} (C^{-1}\hs)_{\ell_2 m_2} (C^{-1}\ha)_{\ell_3 m_3} \right\rangle_{\hs}  \nn
\ea
where $\langle\cdot\rangle_{\hs}$ denotes an average over signal+noise simulations $\hs$.

The normalizing constant ${\mathcal N}$ appearing in our pipeline is computed using non-Gaussian simulations
as described in \cite{Smith:2006ud}, where some computational speedups for the cubic and linear terms are also presented.
This end-to-end normalization ensures that the estimator is unbiased (i.e. $\langle \hfnl \rangle = \fnlloc$)
without making any approximations.

We compute the error $\sigma(\fnlloc)$ by Monte Carlo.  In each simulation, we randomly generate a CMB realization $a$
and a noise realization in each channel, then process the simulation in the same way as the real data.
Thus in our pipeline, the same set of Monte Carlo simulations is used to compute three quantities:
the linear term in the estimator, the normalization ${\mathcal N}$, and the error $\sigma(\fnlloc)$.

When we report estimates of $\fnlloc$ using the suboptimal estimator (typically for purposes of
direct comparison with \cite{Komatsu:2008hk,Yadav:2007yy}), we define the suboptimal estimator by replacing the $C^{-1}\ha$ in 
Eq.~(\ref{eq:Edef3}) by a heuristically-constructed map which is linear in the WMAP data and approximates $C^{-1}\ha$.
This requires making arbitrary choices in several places (e.g. relative weighting of different channels)
and so two implmentations of ``the'' suboptimal estimator will give different results.
We have attempted to follow \cite{Komatsu:2008hk} as closely as possible.
More precisely, our suboptimal estimator is obtained by replacing $C^{-1}\ha$ in Eq.~(\ref{eq:Edef3}) by the quantity
denoted by $a_{\ell m}$ in Eq.~(A27) of \cite{Komatsu:2008hk}.

In principle, our estimator has some nonzero response to the ``equilateral'' three-point signal $\fnleq$
introduced in \cite{Creminelli:2005hu}, and to various three-point correlations between late-universe anisotropies such as
ISW, point sources, gravitational lensing, and thermal SZ.
This nonzero response can be corrected by jointly estimating $\fnlloc$ in combination with additional three-point
signals ($\fnleq, f_{NL}^{\rm ps}, \ldots$), but we do not do so here, and simply use Eq.~(\ref{eq:Edef2})
directly.
This makes it straightforward to compare results with \cite{Komatsu:2008hk,Yadav:2007yy}, where the same approach was used.
Including joint estimation of $\fnleq$ should not appreciably change the results, since
the cross-correlation between the local and equilateral shapes is small, and the equilateral shape
is not detected in WMAP.
Some secondary contributions to $\hfnl$ have been studied and have all been predicted to be small compared to the WMAP statistical error
\cite{Smith:2006ud,Seljak:1998nu,Cooray:1999kg,Goldberg:1999xm,Verde:2002mu,Castro:2002df,Babich:2008uw,Senatore:2008wk},
although it is not clear that all possible secondaries have been studied.

Throughout this paper, we have shown the $\ell$-dependence of our $\fnlloc$ estimates in two ways: either
by plotting a cumulative $\fnlloc$ estimate versus $\ellmax$ (e.g. top panel of Fig.~\ref{oldvsoptimal}), 
or by defining bins in $\ell$ and reporting an independent estimate of $\fnlloc$ in each bin (e.g. bottom panel of Fig.~\ref{oldvsoptimal}).
In the second case, the $\fnlloc$ estimate for a bin $[\ell_0,\ell_1]$ is defined by restricting the
sum in the estimator (Eq.~(\ref{eq:Edef2})) to triples $(\ell_1,\ell_2,\ell_3)$ which satisfy
$\ell_0 < \max(\ell_1,\ell_2,\ell_3) \le \ell_1$.
In implementation, it is convenient to note that the binned and unbinned estimators are related by:
\be
\hfnl(\ell_0,\ell_1) = \frac{{\mathcal N}_1 \hfnl(\ellmax=\ell_1) - {\mathcal N}_0 \hfnl(\ellmax=\ell_0)}{{\mathcal N}_1 - {\mathcal N}_0}
\ee
where $\hfnl(\ell_0,\ell_1)$ denotes the binned estimator and $\hfnl(\ellmax=\ell_i)$ denotes the unbinned
estimator with normalization constant ${\mathcal N}_i$.  (Note that the normalization constant ${\mathcal N}$ appearing in Eq.~(\ref{eq:Edef2})
depends on $\ellmax$.)

\end{document}